\documentclass{svmult}
\usepackage[english]{babel}
\usepackage{amsmath,amssymb,amscd,graphicx,bbm}

\usepackage{makeidx}         
\usepackage{graphicx}        
\usepackage{multicol}        
\usepackage[bottom]{footmisc}

\makeindex             

\catcode`\>=\active \def>{
\fontencoding{T1}\selectfont\symbol{62}\fontencoding{\encodingdefault}}
\catcode`\|=\active \def|{
  \fontencoding{T1}\selectfont\symbol{124}\fontencoding{\encodingdefault}}
\newcommand{\bignone}{}
\newcommand{\mathd}{\mathrm{d}}
\newcommand{\mathe}{\mathrm{e}}

\newcommand{\tmem}[1]{{\em #1\/}}
\newcommand{\tmop}[1]{\ensuremath{\operatorname{#1}}}
\newcommand{\tmtextbf}[1]{{\bfseries{#1}}}

\begin{document}

\title*{Non-Markov processes in quantum mechanics}
\author{Bassano Vacchini}
\institute{Dipartimento di Fisica ``Aldo Pontremoli'', Universit\`a degli Studi di Milano
\\
Via Celoria 16, Milan, 20133, Italy\\
INFN Sezione di Milano\\ Via Celoria 16, Milan, 20133, Italy
\\
\texttt{bassano.vacchini@mi.infn.it}}
%
%

\title{Non-Markovian processes in quantum theory}

\maketitle

\begin{abstract}
  The study of quantum dynamics featuring memory effects has always been a
  topic of interest within the theory of open quantum system, which is
  concerned about providing useful conceptual and theoretical tools for the
  description of the reduced dynamics of a system interacting with an external
  environment. Definitions of non-Markovian processes have been introduced
  trying to capture the notion of memory effect by studying features of the
  quantum dynamical map providing the evolution of the system states, or
  changes in the distinguishability of the system states themselves. We
  introduce basic notions in the framework of open quantum systems, stressing
  in particular analogies and differences with models used for introducing
  modifications of quantum mechanics which should help in dealing with the
  measurement problem. We further discuss recent developments in the treatment
  of non-Markovian processes and their role in considering more general
  modifications of quantum mechanics.
\end{abstract}

\section{Introduction}

Quantum theory was born as a new mechanics, capable of providing the correct
quantitative assessment of phenomena which could not find their explanation
within the usual framework of classical mechanics. About a century after its
introduction, many different facets and complementary presentations of the
theory have been worked out, putting into evidence in particular that quantum
theory indeed provides a new probabilistic framework for the prediction of
outcomes of statistical experiments. It is therefore not only a ``quantum''
version of classical mechanics, it is indeed a \ ``quantum'' version of
classical probability theory, containing into itself an often non trivial
classical limit {\cite{Streater2000a,Strocchi2005,Vacchini2010a}}. One of the
most intriguing and delicate aspects of quantum theory is its irreducibly
probabilistic structure, conflicting with the deterministic description we are
accustomed to, as well as our everyday experience of the realization of
definite events. From a classical viewpoint a probabilistic analysis is only
necessary if not all degrees of freedom are under control or can be taken into
account in detail. Not so for quantum theory. This state of affairs has led
among others to the so-called ``measurement problem'', referring to the
difficulty in reconciling the classical description for macroscopic objects
and the laws of quantum theory, predicting a statistical distribution rather
than definite events {\cite{Schlosshauer2005a}}. On turn, this problem has led
to consider alternatives to quantum theory, complying with its successes but
leading to a different behavior for the prediction of events, effectively
suppressing superposition of macroscopic objects. Among these theories one of
the most renowned classes is given by collapse model, also known as dynamical
reduction models {\cite{Bassi2003a,Bassi2013a}}, arisen from the seminal paper
{\cite{Ghirardi1986a}}. Their distinctive trait is a stochastic non-linear
modification of the Schr{\"o}dinger equation, which on top of the standard
evolution allows for the introduction of a collapse or localization mechanism.
This mechanism, once accepted, avoids the measurement problem. Importantly,
this mechanism has to be implemented at the level of the wavefunction, so as
to allow for the suppression of superpositions. Nevertheless, at the level of
experimental observations, it usually cannot be distinguished from other
effects leading to a vanishing contribution of coherences.

The theory of open quantum system is focussed on the description of
the reduced dynamics of a system interacting with other degrees of
freedom, typically called environment, which are not described in
detail {\cite{Breuer2002,Rivas2012}}. The environment therefore brings
in an additional level of randomicity in the dynamics, on top of the
unavoidable statistical aspect brought in by quantum theory.  In this
framework the suppression of superposition states in a given basis is
indeed predicted for a class of models known as decoherence models
{\cite{Hornberger2009a}}.  It thus appears that such models,
bringing in another element of probabilistic description, typically
provide the same average effect as dynamical reduction models, aimed
at overcoming the inherent statistical structure of any quantum
dynamics. In this respect, the two fields of dynamical reduction model
and open quantum system share some underlying mathematical structure,
and we will briefly address recent advancements in open quantum system
having this perspective in mind. An important caveat to be mentioned
is the fact that decoherence models do not provide a solution of the
measurement problem in the sense addressed by collapse models, since
the suppression of macroscopic superpositions only takes place in the
average and a whole statistical distribution of outcomes is predicted
{\cite{Joos2003}}.

The contribution is organized as follows. In Sect.~2 we briefly outline the
open quantum system viewpoint and address the term quantum process as used in
the physical literautre. The description of decoherence effects and their
relationship to specific collapse models is worked out in Sect.~3. Finally
Sect.~4 is devoted to introduce the notion of non-Markovian dynamics for an
open system, and its influence on the elaboration of dynamical reduction
models.

\section{Open systems and quantum processes}

\begin{figure}
 \includegraphics[width=0.99\textwidth]{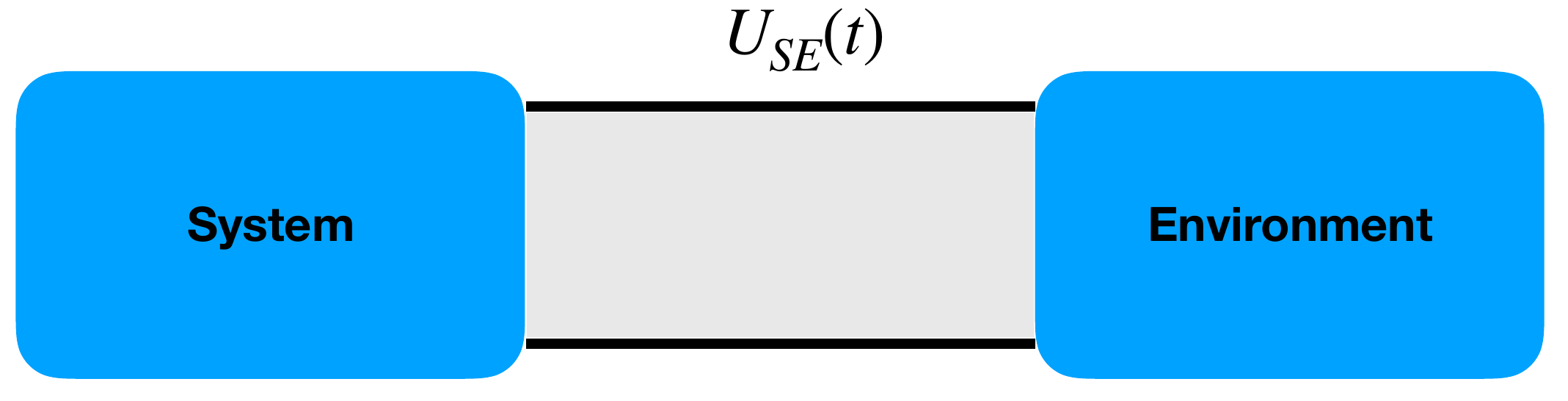}
\caption{Illustration of an open system interacting with an environment
  via a unitary coupling $U_{S E} ( t )$. \label{fig:oqs0}}
\end{figure}

For the case in which a quantum system is not isolated from other quantum
systems, the latter should be taken into account in the description of its
dynamics. If the system and the other degrees of freedom, collectively named
environment, do not share correlations at the initial time, one can describe
the evolution of the system alone by introducing a collection of completely
positive trace preserving maps $\{ \Phi ( t ) \}_{t \in \mathbbm{R}_{+}}$,
which determine the statistics of any local observation once the initial state
of the system $\rho_{S} ( 0 )$ has been specified according to the formula
\begin{eqnarray}
  \langle A_{S} \rangle_{t} & = & \tmop{Tr} \{ A_{S} \Phi ( t ) [ \rho_{S} ( 0
  ) ] \} , \nonumber
\end{eqnarray}
where $A_{S}$ denotes a system observable. The collection of maps $\{ \Phi ( t
) \}_{t \in \mathbbm{R}_{+}}$ describes what is usually called a quantum
process. The term process is here used in a loose sense, in analogy with the
classical situation, hinting at the presence of an irreducible randomicity,
here corresponding to the environmental degrees of freedom not accessible or
described in detail, but affecting the system dynamics due to a
unitary coupling with the environment $U_{S E} ( t )$ as drawn in Fig.~\ref{fig:oqs0}.
 If system and environment interaction can be neglected,
and only in this case, $\Phi ( t )$ is a unitary transformation, implying in
particular a group composition law. In all other cases irreversibility is
lost, and the general mathematical structure of this collection of maps is not
known. Some partial results are however available. A most famous and relevant
class of reduced dynamics is obtained if we ask $\Phi ( t )$ to obey a
semigroup composition law forward in time. For this case we have $\Phi ( t ) =
\exp ( t\mathcal{L} )$, with $\mathcal{L}$ in Lindblad form
{\cite{Breuer2002}}, that is
\begin{eqnarray}
  \mathcal{L} [ \rho_{S} ( t ) ] & = & - \frac{i}{\hbar} [ H, \rho_{S} ( t ) ]
  + \sum_{k} \lambda_{k} \left[ A_{k} \rho_{S} ( t ) A_{k}^{\dag} -
  \frac{1}{2} \{ A_{k}^{\dag} A_{k} , \rho_{S} ( t ) \} \right] \bignone ,
  \nonumber
\end{eqnarray}
where $\{ A_{k} \}$ and $H$ denote system operators, with \ $H$ an effective
self-adjoint Hamiltonian. A dynamics of this kind has always been called
Markovian, since it arose as quantum counterpart of classical Markovian
semigroups. The implicit idea is that the stochasticity in the dynamics
arising due to interaction with the environment does not lead to effects that
can be termed memory, making reference to previous history or states of the
system. This feature is immediately lost even only considering dynamics which
can be obtained as random mixture of unitary evolutions, so-called random
unitary dynamics
{\cite{Audenaert2008a,Pernice2012a,Vacchini2012a,Chruscinski2013a}}, which
might arise also as a consequence of classical environment noise and can be
experimentally engineered {\cite{Cialdi2017a,Rossi2017a}}. The operators $\{
A_{k} \}$ describe microscopic interaction events, e.g. random localization or
momentum transfer events for the case of decoherence as discussed in \
Sect.~3.

\section{Events and decoherence}

Dynamical reduction models and open quantum system theory share a common root
in the treatment of measurement in quantum mechanics, to be seen as dealing
with a description of the outcomes of statistical experiments in which the
interaction with the measurement apparatus is taken into account. Indeed, the
first seminal contributions to open quantum systems were intimately connected
with the description of measurement processes, and its relevance for the
foundations of quantum mechanics {\cite{Davies1976,Ludwig1983,Kraus1983}},
putting in particular into evidence the relevance of the mathematical notion
of complete positivity. Not by chance the original GRW paper, which introduced
the first collapse model, was built upon work aimed at the quantum description
of continuous measurement in time {\cite{Barchielli1982a,Barchielli1983a}},
and started the treatment from a master equation describing decoherence in
position {\cite{Vacchini2007b}}.

To better work out this connection, let us consider in more detail how a
collapse model can describe in the average a decoherence effect and how a
microscopic description of decoherence can be related to a notion of event. In
this spirit we briefly recall the formulation of the GRW model in the
formulation via stochastic differential equations
{\cite{Bassi2003a,Smirne2014a}}.
\begin{equation}
  \label{eq:edsnl} \mathd | \psi (t) \rangle =- \frac{i}{\hbar}  \hat{H}_{0}
  \psi (t) \mathd t+ \int_{\mathbbm{R}} \mathd y \left( \frac{L(y, \hat{x}
  )}{\|L(y, \hat{x} )| \psi (t) \rangle \|} - \mathbbm{1} \right) | \psi (t)
  \rangle \mathd N (y,t) ,
\end{equation}
where $\psi (t)$ is the system's wavefunction, $\hat{H}_{0}$ denotes the
Hamiltonian appearing in the standard Schr{\"o}dinger equation and the
stochastic modification is determined by the collection of operators $\{ L(y,
\hat{x} ) \}_{y \in \mathbbm{R}}$, with $\hat{x}$ the standard position
operator, and the family of classical stochastic processes $\{ N(y,t) \}_{y
\in \mathbbm{R}}$. Note in particular that this modification is non-linear. In
order to obtain suppression of spatial superposition of states, the $L$
operators have to act as localization operators and to recover the original
GRW model must be of the form
\begin{equation}
  \label{eq:locop} L (y, \hat{x} ) = \frac{1}{\sqrt[4]{\pi r_{c}}} e^{-
  \frac{(y- \hat{x} )^{2}}{2r^{2}_{c}}} .
\end{equation}
The stochastic modification depends on the field of independent processes $\{
N(y,t) \}_{y \in \mathbbm{R}}$ such that $N (y,t) dy$ is the counting process
giving the number of jumps taking place at time $t$ in the space interval from
$y$ to $y+dy$. The collection of counting processes satisfies $\mathd N (x,t)
\mathd N (y,t) = \delta (x-y) \mathd N (y,t)$, with rates given by
\[ \begin{array}{lll}
     \mathbbm{E} [ \mathd N(y,t)] & = & \lambda \|L(y, \hat{x} )| \psi (t)
     \rangle \|^{2} \mathd t.
   \end{array} \]
The phenomenological parameters $\lambda$ and $r_{c}$ determine intensity and
localization strength of the random jumps inducing a dynamical localization in
position of the system. Averaging over the realization of the processes one
obtains the state determining the statistics of observation on the system,
namely
\begin{eqnarray}
  \rho ( t ) & = & \mathbbm{E} [ | \psi (t) \rangle \langle \psi ( t ) | ] ,
  \nonumber
\end{eqnarray}
which obeys the master equation
\begin{eqnarray}
  \label{eq:ms} \frac{\mathd}{\tmop{dt}} \rho ( t ) & = & - \lambda \left[
  \rho ( t ) - \int dyL(y, \hat{x} ) \rho ( t ) L(y, \hat{x} ) \bignone
  \right] 
\end{eqnarray}
predicting a reduction of the off-diagonal matrix elements in the position
representation according to
\begin{eqnarray}
  \label{eq:decoh} \langle x| \rho ( t ) |y \rangle & = & \exp \left( -
  \lambda t \left[ 1- \int dz L (z,x) L (z,y) \right] \right) \langle x| \rho ( 0 )
  |y \rangle . 
\end{eqnarray}
The obtained master Eq.~(\ref{eq:ms}) is in standard Lindblad form
{\cite{Breuer2002}}, describes decoherence in position according to
Eq.~(\ref{eq:decoh}), and in particular is characterised by translational
invariance. Building on this aspect one realizes that it can be written in an
explicit translationally covariant form
{\cite{Holevo1993a,Vacchini2001b,Vacchini2005a}} as follows
\begin{eqnarray}
  \label{eq:ms2} \frac{\mathd}{\tmop{dt}} \rho ( t ) & = & - \lambda \left[
  \rho ( t ) - \int dq \tilde{L} (q) \mathe^{\frac{i}{\hbar} q \hat{x}} \rho (
  t ) \mathe^{- \frac{i}{\hbar} q \hat{x}} \bignone \right] 
\end{eqnarray}
with $\tilde{L} (q)$ Fourier transform of the function $L^{2} (y,0)$, that is
again a Gaussian weight. It thus appears that the dynamics that can be
observed as a consequence of the localization mechanism, described at the
level of trajectories of the wavefunction in Hilbert space by the stochastic
differential equation Eq.~(\ref{eq:edsnl}), is the same that would arise as a
consequence of interaction of the system with an external environment whose
effect can be described in terms of localisation events as in
Eq.~(\ref{eq:ms}) or in terms of momentum transfers described by the
collection of unitaries $\left\{ \mathe^{\frac{i}{\hbar} q \hat{x}}
\right\}_{q \in \mathbbm{R}}$ as in Eq.~(\ref{eq:ms2}). This viewpoint,
connecting the open system based description of decoherence and the
measurement based viewpoint of collapse models, implies in particular that the
natural benchmark in the assessment of possible modifications of the quantum
mechanical predictions due to a collapse mechanism is the estimate of possible
decoherence effects affecting the considered dynamics. Indeed, this is one of
the main difficulties in looking for experimental signatures of collapse
mechanisms {\cite{Bassi2013a}}. On the other hand awareness of this
relationship has opened the way to consider variants of dynamical reduction
models. In particular, it has led to overcome an important intrinsic
limitation of models such as Eq.~(\ref{eq:edsnl}), which predict an infinite
growth of the system energy {\cite{Bassi2005b,Smirne2014a}}. A further natural
extension of dynamical reduction model arising from analogy and differences
shared with open quantum system models is the inclusion of memory effects
{\cite{Bassi2009a,Bassi2009b,Ferialdi2012a,Ferialdi2012b,Ferialdi2017b}}, in
view of a definition of non-Markovian dynamics as discussed in Sect.~4.

\section{Non-Markovian processes}

In mentioning some of the basic tenets and results of the theory of open
quantum systems, we have put into evidence the notion of quantum process as
used and understood in the physical literature. In particular, the time
evolutions arising as solutions of master equations in Lindblad form are
typically termed quantum Markovian processes, since they provide the natural
quantum counterpart of classical semigroup evolutions, arising in connection
with homogeneous in time Markovian processes. A next natural step in this
respect is considering time evolutions which can provide a quantum realization
of a non-Markovian process. Given the looser definition of process considered
in the quantum framework, as a collection of time dependent completely
positive trace preserving maps describing a continuous quantum dynamics, one
might consider a suitable definition of non-Markovian quantum process within
this very same framework of dynamical maps. Indeed, providing a notion of
non-Markovian quantum process in the same spirit as in the classical case,
which gives an exact defintion of Markovian process in terms of conditions on
the infinite hierarchy of conditional probability densities for the process,
appears to be a very difficult task. Already from a conceptual point of view
the situation does not appear to be neatly defined, since speaking about
values of an observable at a given time calls for a measurement procedure
which affects the subsequent values to be assumed by the quantity
{\cite{Vacchini2011a}}. On the contrary, focusing on the collection of
completely positive trace preserving maps giving the reduced dynamics has
allowed to introduce clearcut definitions of Markovian, and in a complementary
way non-Markovian, quantum process. Actually, there have been various
proposals in this direction. We will here only focus on one of them, based on
the behavior of the distinguishability of states in time, which is in direct
relationship with a notion of divisibility of the time evolution maps. For
more details and a complete treatment we refer the reader to recent reviews
{\cite{Breuer2012a,Rivas2014a,Breuer2016a,Devega2017a}}.

\begin{figure}
 \includegraphics[width=0.99\textwidth]{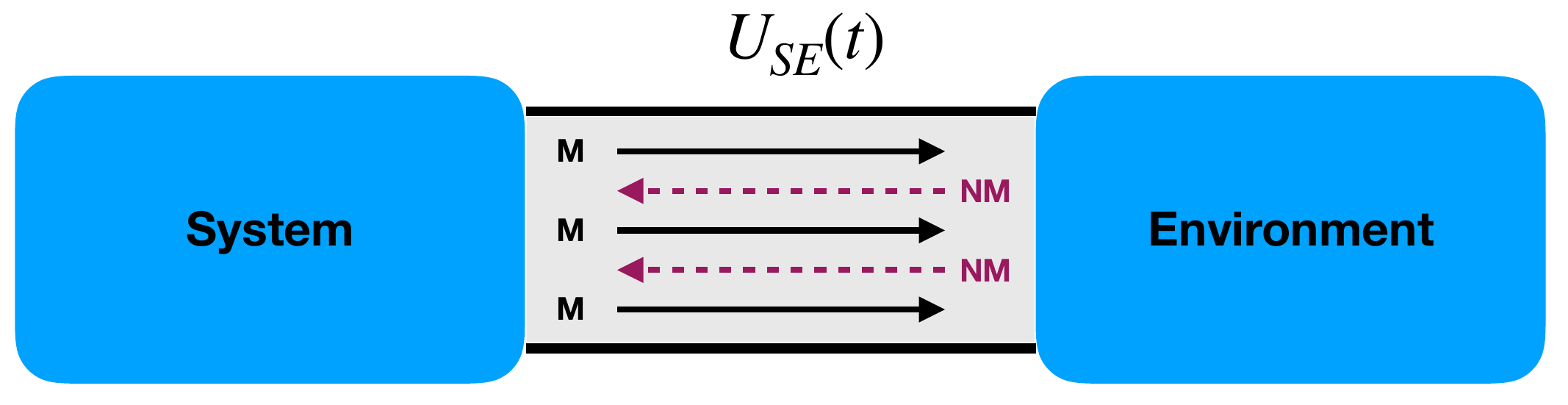}
\caption{Open system interacting with an environment
  via a unitary coupling $U_{S E} ( t )$. Markovian effects (M) are depicted
  as an information flow from system to environment, while an information flow
  from environment to system (NM) is identified with memory
  effects.\label{fig:oqs}}
\end{figure}

The basic insight can be summarized as follows. By interacting with the
environmental degrees of freedom the system gets correlated with the
environment and possibly leads to a change in time of the reduced state of the
environment itself. As a consequence of the dynamics therefore, the capability
of distinguishing two different initial system states, by performing
measurements on the system degrees of freedom only, changes in time. Indeed,
taking the partial trace necessary to define the reduced system state, which
is all that is necessary in order to provide the statistics of measurements on
the system, the whole information about correlations is no more available. To
exploit this fact one can introduce a suitable quantifier of the
distinguishability between states, such as the trace distance, given by the
trace norm of the difference of the states
\begin{eqnarray}
  \label{eq:td} D ( \rho_{S}^{1} ( t ) , \rho_{S}^{2} ( t ) ) & = &
  \frac{1}{2} \| \rho_{S}^{1} ( t ) - \rho_{S}^{2} ( t ) \|_{1} 
\end{eqnarray}
and consider its behavior in time. Being a contraction under the action of
completely positive trace preserving transformations, the trace distance
always diminishes with respect to its initial value, that is
\begin{eqnarray}
  \label{eq:tdto} D ( \rho_{S}^{1} ( t ) , \rho_{S}^{2} ( t ) ) & \leqslant &
  D ( \rho_{S}^{1} ( 0 ) , \rho_{S}^{2} ( 0 ) ) . \nonumber
\end{eqnarray}
In particular for the semigroup case, considered in Sect.~2 for the case of a
quantum Markovian process, due to the composition law one has a monotonous
reduction of the distance among states with time. In such a situation the
distance between states, and therefore their distinguishability
{\cite{Fuchs1999a}}, gets smaller and smaller with elapsing time. The failure
of this monotic decreasing behavior for at least a couple of possible initial
states has been taken as indication of non-Markovian dynamics in the seminal
paper {\cite{Breuer2009b}}, since it amounts to a revival in the
distinguishability between the states that can only arise as a consequence of
previously established correlations with the environment or changes in the
environmental state that affect the subsequent reduced system dynamics. This
fact is schematically drawn in Fig.~\ref{fig:oqs}.
The validity of this interpretation is substantiated by the inequality
{\cite{Laine2010b,Breuer2018a,Campbell2019b}}
\begin{eqnarray}
  D ( \rho^{1}_{S} ( t ) , \rho^{2}_{S} ( t ) ) -D ( \rho^{1}_{S} ( s ) ,
  \rho^{2}_{S} ( s ) ) & \leqslant & D ( \rho^{1}_{SE} ( s ) , \rho^{1}_{S} (
  s ) \otimes \rho^{1}_{E} ( s ) )  \label{eq:bound}\\
  &  & +D ( \rho^{2}_{SE} ( s ) , \rho^{2}_{S} ( s ) \otimes \rho^{2}_{E} ( s
  ) ) \nonumber\\
  &  & +D ( \rho^{1}_{E} ( s ) , \rho^{2}_{E} ( s ) ) , \nonumber
\end{eqnarray}
where it is assumed that $t \geqslant s$. The term at the lhs when positive
provides a signature of non-Markovianity, so that the positivity of the rhs is
a precondition for non-Markovianity, to be traced back to the effects
mentioned above: correlations and influence of the system on the environment.
While the notions of distinguishability, contractivity of the used
distinguishability quantifier upon the action of a quantum transformation, and
connection of the distinguishability revivals to the imprint of the system
dynamics left in correlations or environment, provide the basic traits of this
approach to the description of memory effects in quantum mechanics, many more
subtle issues are involved in the definition of this framework. Importantly,
there is a stringent mathematical connection between this viewpoint and
divisibility properties of the time evolution, corresponding to the fact that
the evolution over a finite time can always be split into evolutions over
shorter times, each described by a proper quantum transformation
{\cite{Rivas2010a,Chruscinski2011a,Wissmann2015a}}.

Dynamics allowing for non-Markovian effects have also been considered in the
above-mentioned framework of a decoherence dynamics driven by random events
{\cite{Vacchini2008a,Smirne2010a}}, as well as in the introduction of more
general dynamical reduction models {\cite{Adler2007a,Ferialdi2012a}}. While in
the context of decoherence allowing for non-Markovian dynamics is a way to
consider more general and accurate description of the reduced dynamics, within
the framework of \ dynamical reduction models non-Markovian models lead to
possibly more stringent exclusion regions of the parameter values which
characterise the model.

\section{Conclusions and Outlook}

In recent times a lot of work in the field of open quantum system has been
devoted to characterization and study of non-Markovian dynamics. This research
has involved both the very definition and clarification of what can be meant
as quantum dynamics featuring memory effects, as well as the possible
relevance of non-Markovian dynamics in the description of the reduced dynamics
of non isolated quantum systems as well as related fields. In this
contribution we have recalled in particular the relationship between the
description of decoherence in open quantum system and modifications of quantum
mechanics such as dynamical reduction models introduced for the sake of better
grasping the so-called quantum measurement problem. We have briefly discussed a
natural physical interpretation of non-Markovian dynamics as related to
information exchange between system and environment, and pointed to the use of
the formalism of non-Markovian dynamics to consider more general collapse
model which might help in improving the known bounds on the parameters
characterizing the possible deviations from standard quantum mechanics. The
relevance of the classification of non-Markovian dynamics itself as well as
the role of memory effects in collapse mechanisms remain two open questions
that will surely involve future research.

\section*{Acknowledgements}

The author acknowledges support from the Joint Project ``Quantum Information
Processing in Non-Markovian Quantum Complex Systems'' funded by FRIAS,
University of Freiburg and IAR, Nagoya University, from the FFABR project of
MIUR and from the Unimi Transition Grant H2020.

\end{document}